\documentclass{article}
\usepackage{spconf,amsmath,graphicx,hyperref,caption}
\title{WordVoice: Explicit and Decoupled Multi-Dimensional \\ Word-Level Control for LLM-Based TTS}

\name{
    Sihang Nie$^{1,2}$\sthanks{Work conducted when the author was intern at Huya Inc.}, 
    Jinxin Ji$^{3,4}$,
    Xiaofen Xing$^{1}$\sthanks{Corresponding author.}, 
    Deyi Tuo$^{2}$,
    Chengbin Jin$^{2}$,
    Jialong Mai$^{1}$,
    Xiangmin Xu$^{1,5}$
}

\address{
  $^{1}$South China University of Technology,
  $^{2}$Huya Inc.,
  $^{3}$Tongji University \\
  $^{4}$The Hongkong polytechnic university,
  $^{5}$Foshan University \\
  xfxing@scut.edu.cn, bcshnie@mail.scut.edu.cn
}

\usepackage{booktabs}
\usepackage{multirow}
\begin{document}
\maketitle

\begin{abstract}
While recent Large Language Model (LLM)-based Text-to-Speech (TTS) systems have achieved remarkable naturalness, they predominantly rely on implicit end-to-end generation paradigms, resulting in coarse-grained control. In scenarios demanding precise stylistic interventions and strict temporal alignment, such as audiobook narration and video dubbing, the inability to explicitly manipulate word-level acoustic attributes remains a critical bottleneck. This limitation is primarily amplified by the severe scarcity of fine-grained annotated datasets and the architectural challenge of integrating multi-dimensional control signals into discrete autoregressive generation. To address this, we propose a unified framework for highly precise word-level control. First, we construct WordVoice-5A, a massive 4.7k-hour bilingual dataset featuring five-dimensional word-level annotations (duration, boundary, energy, pitch and tone) developed through a rigorous linguistically-guided pipeline. Second, we introduce WordVoice to transform the implicit generation process into an explicit, highly controllable paradigm. Specifically, we introduce a bound-token mechanism within the LLM to formulate an explicit ``acoustic planning'' process, enabling adaptive multi-task prosodic planning and flexible manual intervention. Furthermore, we augment the token-to-waveform stage with a fine-grained acoustic modulation module, bridging the resolution gap to strictly align word-level attributes between highly compressed discrete tokens and continuous waveforms. Extensive experiments demonstrate that WordVoice achieves superior, decoupled control over multiple acoustic dimensions while maintaining competitive zero-shot synthesis stability. The code and audio samples are publicly available at~\url{https://xxh333.github.io/wordvoice-demo/}.
\end{abstract}


\section{Introduction} 
\label{sec:intro}
Recent advancements in controllable Text-to-Speech (TTS) have enabled expressive style manipulation. Existing approaches primarily fall into two paradigms. The first relies on global embedding control~\cite{emosphere++, facespeak, accenttts}, which achieves satisfactory results in specific domains but suffers from coarse control granularity and limited generalization. The second paradigm, instruction-based TTS~\cite{instructtts, cosyvoice2, hdppt}, leverages natural language prompts to enable localized and more generalized control. However, despite their versatility, these models still fall fundamentally short of achieving high-precision, fine-grained acoustic control at the character- or word-level.

\begin{figure}[t]
\begin{minipage}[hbtp]{1.0\linewidth}
  \centering
  \centerline{\includegraphics[width=\linewidth]{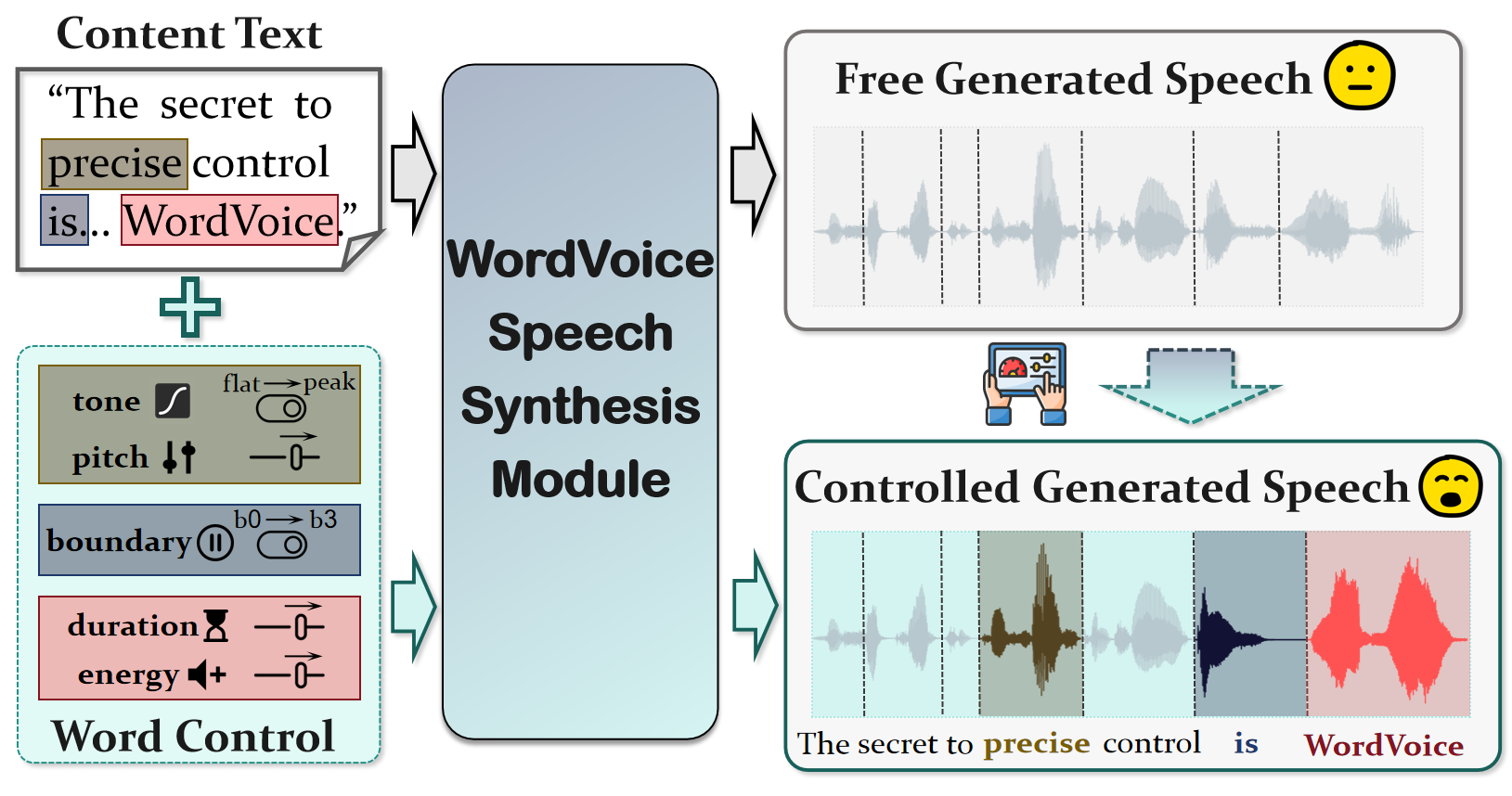}}
\end{minipage}
  \caption{WordVoice framework. By introducing explicit word-level control, WordVoice supports a dual-mode synthesis paradigm. Users can either rely on the model's autonomous prosodic planning or explicitly manipulate five-dimensional acoustic attributes for specific words to achieve highly expressive and precise stylistic interventions.}
  \label{fig:mot}
\end{figure}

This granularity bottleneck creates a critical void in real-world applications that demand deterministic prosodic interventions. For instance, in audiobook narration, creators often need to meticulously manipulate the duration, energy (emphasis), or tone (intonation) of specific words to convey nuanced character emotions~\cite{deepdub, magictts}. Yet, relying solely on textual instructions makes it difficult for current models to reliably enforce such localized, high-precision adjustments. Similarly, in video dubbing, the inability to explicitly dictate exact temporal alignment severely hinders strict cross-modal alignment between speech and lip-sync~\cite{synctalk, diflowdubber}.

\begin{table}[t]
\centering
\caption{Comparison of WordVoice with existing controllable TTS paradigms.}
\label{tab:feature_comparison}
\resizebox{\columnwidth}{!}{%
\begin{tabular}{llcl}
\toprule
\textbf{Method} & \textbf{Control Paradigm} & \textbf{Granularity} & \textbf{Control Dimensions} \\
\midrule
Instruct-based TTS & Implicit (Text Prompt) & Utterance & Global Style, Emotion \\
WeSCon & Implicit (Emotion Label) & Word & Localized Emotion \\
MagicTTS & Explicit (Acoustic) & Word & Duration, Pause \\
\midrule
\textbf{WordVoice (Ours)} & \textbf{Explicit (Acoustic)} & \textbf{Word} & \textbf{Dur, Bnd, Eng, Pit, Ton} \\
\bottomrule
\end{tabular}%
}
\end{table}

Fundamentally, this limitation stems from the architectural evolution of TTS systems. Early TTS models~\cite{fs2, vits} relied heavily on the explicit prediction of acoustic attributes. In contrast, current mainstream TTS paradigms favor end-to-end generation, directly mapping text to the target acoustic space. Since explicit intermediate acoustic representations are absent in this process, these models inherently lack the mechanisms for explicit multi-dimensional control during generation. A few pioneering studies have attempted to reintroduce word-level control, such as WeSCon~\cite{wescon} for localized emotion and MagicTTS~\cite{magictts} for duration adjustments. However, these approaches remain highly unidimensional. A comprehensive framework capable of simultaneously controlling multiple acoustic dimensions at the word-level remains largely unexplored. This research gap is primarily aggravated by the severe scarcity of large-scale, word-level annotated datasets.

To bridge this gap, we propose a unified framework from both data and architectural perspectives. First, by developing a highly accurate, linguistically-guided annotation pipeline, we construct \textbf{WordVoice-5A}, a massive 4.7k-hour bilingual (Chinese and English) dataset featuring five dimensions of word-level acoustic attributes (duration, boundary, energy, pitch, and tone). Building upon this data foundation and the LLM-based TTS paradigm, we introduce \textbf{WordVoice}, as illustrated in Fig.~\ref{fig:mot}. Specifically, we pioneer an ``acoustic planning'' process within the autoregressive LLM via an explicit bound-token mechanism. Rather than passively predicting attributes in parallel, this mechanism restructures the causal generation process to explicitly reason about prosodic execution before generating speech tokens. Furthermore, to complement the acoustic details lost during discrete token quantization, we introduce a fine-grained word-level style modulation module in the token-to-waveform stage. This enhancement further aligns the word-level acoustic attributes between the discrete semantic tokens and the generated continuous waveforms. As summarized in Table~\ref{tab:feature_comparison}, compared to existing controllable TTS paradigms, WordVoice uniquely achieves explicit, multi-dimensional control at the fine-grained word level.

Our main contributions are as follows:
\begin{itemize}
    \item \textbf{Data Foundation:} We construct and open-source WordVoice-5A, a massive 4.7k-hour dataset featuring five-dimensional word-level annotations, along with its linguistically-guided annotation pipeline.
    \item \textbf{Architectural Innovation:} We propose the WordVoice framework, a joint design of ``acoustic planning'' generation and fine-grained style modulation, achieving explicit, multi-dimensional word-level control in LLM-based TTS.
    \item \textbf{Word-level Control:} Extensive experiments demonstrate that WordVoice achieves highly precise and decoupled word-level control across multiple acoustic dimensions.
\end{itemize}

\begin{figure*}[t]
    \centering
    \includegraphics[width=0.72\textwidth]{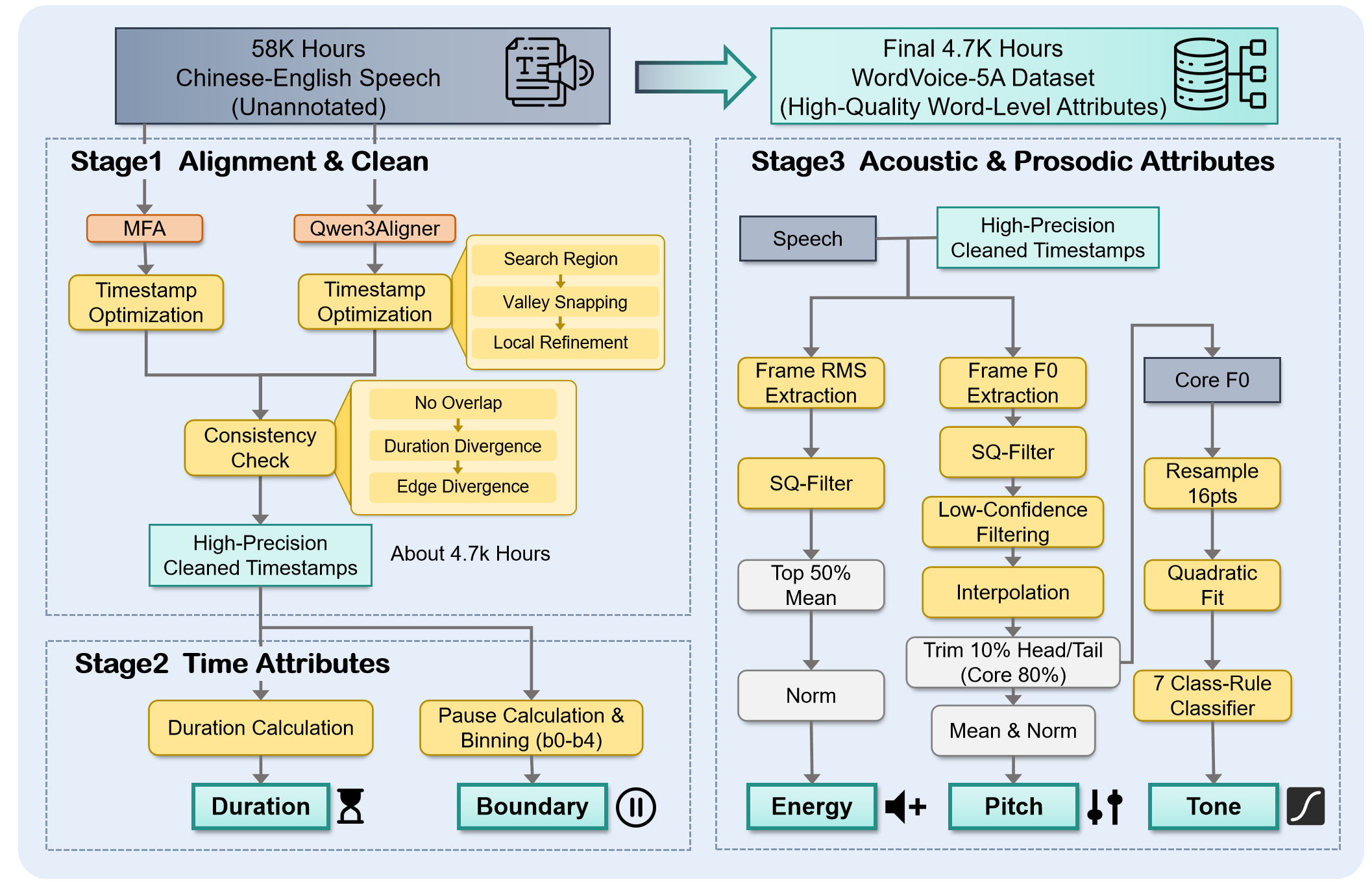} 
    \caption{The linguistically-guided annotation pipeline. (a) Alignment \& Clean: Refining MFA and Qwen3FA timestamps via loudness optimization and consistency checks. (b) Temporal Attributes: Extracting duration and 5-level acoustic boundaries. (c) Acoustic \& Prosodic Attributes: Extracting energy, pitch, and 7-category tone via truncation and morphological modeling.}
    \label{fig:pipeline}
\end{figure*}

\section{Word-Level Annotation Pipeline for WordVoice-5A}
\label{sec:pipeline}

\subsection{Overview of WordVoice-5A}
Several large-scale open-source speech corpora have significantly advanced the development of zero-shot TTS~\cite{emilia, textrolspeech, speechcraft}. However, these datasets are typically limited to utterance-level transcripts and instructions. Although the recent LEMAS dataset~\cite{lemas} provides word-level timestamps, its simplistic pipeline yields suboptimal alignment accuracy for high-fidelity modeling. To address this, we source raw audio and transcripts from LEMAS and re-annotate them using our linguistically-guided pipeline to construct WordVoice-5A. Comprising approximately 4.7k hours of bilingual speech, WordVoice-5A provides rigorous word-level annotations across five dimensions (duration, boundary, energy, pitch, and tone). The dataset statistics are detailed in Table~\ref{tab:dataset}.

\begin{table}[t]
\centering
\caption{Statistical overview of the WordVoice-5A dataset.}
\label{tab:dataset}
\resizebox{0.9\columnwidth}{!}{%
\begin{tabular}{lcc}
\toprule
\textbf{Subset} & \textbf{Duration (h)} & \textbf{Characters/Words} \\
\midrule
WordVoice-5A-zh & $\sim$2546 & $\sim$33.27M \\
WordVoice-5A-en & $\sim$2138 & $\sim$18.99M \\
\midrule
\textbf{Total} & \textbf{$\sim$4684} & \textbf{$\sim$52.26M} \\
\bottomrule
\end{tabular}%
}
\end{table}

\subsection{Word-Level Annotation Pipeline}
As illustrated in Fig.~\ref{fig:pipeline}, our automated annotation pipeline consists of three main stages: timestamp alignment and cleaning, temporal attribute annotation, and acoustic/prosodic attribute annotation.

\subsubsection{Timestamp Extraction, Optimization, and Consistency Check}
Accurate word-level timestamps serve as the fundamental prerequisite for extracting all subsequent acoustic attributes. To mitigate the alignment noise inherent in automated tools, we first employ a dual-model extraction using Montreal Forced Aligner (MFA)\footnote{\url{https://pypi.org/project/Montreal-Forced-Aligner/}} and Qwen3FA\footnote{\url{https://huggingface.co/Qwen/Qwen3-ForcedAligner-0.6B}}.

Since traditional aligners often falsely incorporate silences or coarticulation segments into word edges, we introduce a loudness-based optimization mechanism guided by phonation characteristics. 
Specifically, we define a valid \textit{search region} restricted to a 10\% shift of adjacent word durations, strictly preventing the adjusted edge from crossing the syllable nucleus (loudness peak). 
Within this region, if frames fall below a predefined voice activity threshold (indicating silence), \textit{Valley Snapping} is applied to anchor the timestamp to the nearest low-loudness frame. 
Conversely, if all frames within the region exceed the threshold, \textit{Local Refinement} is triggered to shift the edge to the local loudness minimum~\cite{nuclei}.

Following the optimization, we perform a rigorous consistency check to cross-validate the refined timestamps from both models. An utterance is entirely discarded if any constituent word triggers one of the following conditions: 
1) \textit{No Overlap}: zero temporal intersection for the same word between the two models; 
2) \textit{Duration Divergence}: a significant discrepancy in the predicted duration of the word;
and 3) \textit{Edge Divergence}: an excessive shift in the predicted start or end edges.
The specific thresholds for these divergences are calibrated to retain only the top 8\% highest-quality aligned data. 
This rigorous filtering ensures that only highly reliable alignments are passed to the subsequent attribute extraction.

\subsubsection{Temporal Attributes: Duration and Boundary}
Temporal attributes form the structural backbone of speech rhythm and pacing. Following the timestamp optimization, the duration is calculated directly from the temporal spans of the aligned words. For acoustic boundaries, rather than treating pauses as continuous silence, we discretize them into five linguistically meaningful levels. This discretization is inspired by prosodic annotation frameworks like ToBI~\cite{tobi} and C-ToBI~\cite{ctobi}, which map acoustic pauses to syntactic and cognitive phrasing. Specifically, the boundaries are categorized as:
\textit{b0} (continuous phonation with no pause), 
\textit{b1} ($\le$0.05s, intra-word coarticulation micro-pauses), 
\textit{b2} ($\le$0.18s, standard word boundaries), 
\textit{b3} ($\le$0.4s, comma-level prosodic boundaries),
and \textit{b4} ($>$0.4s, period-level terminal boundaries). 
This hierarchical classification provides explicit structural priors for the generative model.

\begin{figure*}[t]
    \centering
    \includegraphics[width=0.94\textwidth]{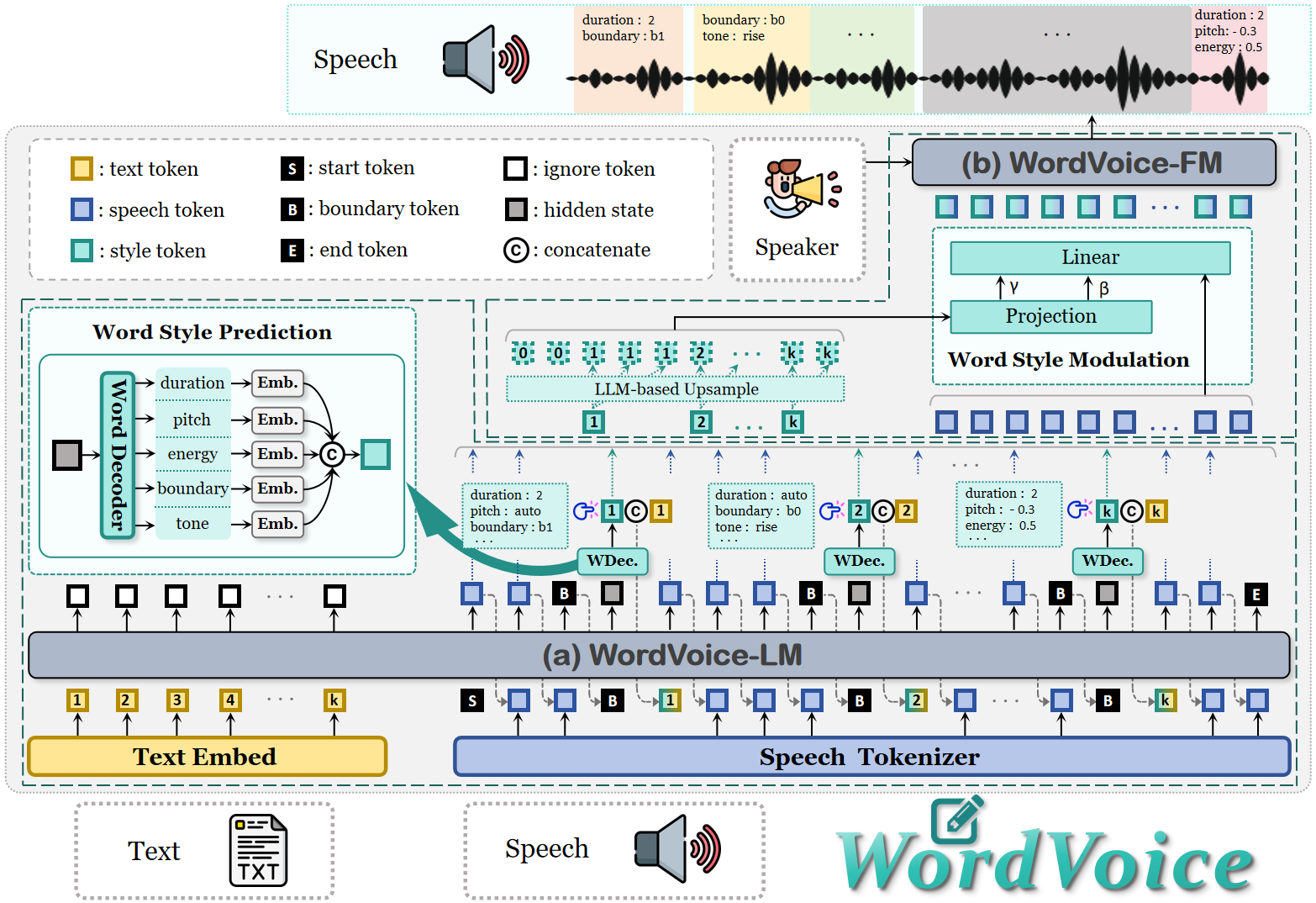}
    \caption{Overall architecture of WordVoice. (a) WordVoice-LLM: During autoregressive decoding, the bound token $\langle b \rangle$ triggers the prediction of five acoustic attributes. These form a word-level style token to explicitly guide the chunked generation of speech tokens. (b) WordVoice-FM: The style tokens are LLM-based upsampled and injected into the Flow Matching backbone, providing fine-grained word style modulation for high-fidelity waveform synthesis.}
    \label{fig:wordvoice}
\end{figure*}

\subsubsection{Acoustic and Prosodic Attributes: Energy, Pitch, and Tone}
To isolate core word attributes from coarticulation, we combine robust signal processing with expert-guided morphological modeling. Initially, frame-level RMS energy and $F_0$ are extracted, smoothed via a Savitzky-Golay filter~\cite{sgfilter}, and log-transformed. Unvoiced or low-confidence $F_0$ frames are masked and interpolated via PCHIP~\cite{pchip}. 

For word-level energy, we average only the top 50\% of frame-level values. This effectively isolates the syllable nucleus loudness, avoiding interference from low-energy boundaries or coarticulatory regions~\cite{nuclei, protrusions}. 
For word-level pitch, we average the central 80\% of the $F_0$ contour. This bilateral truncation aligns with phonetic practices by mitigating onset/coda coarticulation and segmental perturbations~\cite{f0_normalization}. After clipping outliers, energy and pitch are normalized to $[0, 1]$ and $[-1, 1]$ respectively, treating them as core acoustic features for tone production~\cite{mandarin_noise}.

Finally, to establish a unified prosodic representation for both tonal and intonational languages~\cite{promon2009qta}, we propose an expert-guided morphological modeling approach for word-level tone. 
For intonational languages, like English, these morphologies effectively capture word-level pitch accents and local intonation contours. 
The continuous trajectory of the core 80\% $F_0$ is uniformly resampled to 16 points and modeled via quadratic polynomial regression. 
Based on the joint distribution of the fitted curve's curvature, slope, and symmetry axis, decision trees designed by phonetic experts map the continuous contours into seven universal prosodic morphologies: flat, rise, strong rise, fall, strong fall, peak, and valley. Inspired by the discrete, linguistically interpretable labeling traditions of ToBI and C-ToBI~\cite{tobi, ctobi}, this abstraction bridges the gap between continuous acoustic phenomena and symbolic prosodic structures. Consequently, it preserves linguistic interpretability while providing highly controllable, discrete conditions for downstream generation.

\section{Proposed Method: WordVoice}
\label{sec:method}

\subsection{Method Overview}
Mainstream LLM-based TTS models~\cite{indextts2, qwen3tts, mosstts} typically cascade an autoregressive LLM for discrete speech token generation with a non-autoregressive model (e.g., Flow Matching) for waveform synthesis. While yielding impressive zero-shot performance, this paradigm treats speech as an undifferentiated stream, lacking the explicit structural boundaries necessary for precise word-level intervention. 

To address this, we propose WordVoice, as illustrated in Fig.~\ref{fig:wordvoice}. Built upon the CosyVoice3~\cite{cosyvoice3} backbone, our framework refines conventional sentence-level generation into a word-level ``acoustic planning'' and chunked generation process. Instead of blindly predicting total speech tokens, the model explicitly plans each word's acoustic attributes beforehand. This highly controllable, coarse-to-fine framework operates in two stages: 1) a bound-token guided mechanism within the LLM handles macro-prosodic planning, and 2) a fine-grained acoustic modulation module within the Flow Matching stage ensures micro-acoustic fidelity.

\subsection{WordVoice-LLM: Bound-Token Guided Control}
To operationalize this ``acoustic planning'' process, we explicitly introduce a special boundary token $\langle b \rangle$. Unlike conventional multi-task prediction, this mechanism inherently restructures the causal generation process. During decoding, once the model predicts $\langle b \rangle$ for the $i$-th word $w_i$, its current hidden state is immediately routed into a lightweight \textit{Word Decoder} to simultaneously predict the five acoustic attributes $\mathcal{A}_i = \{dur_i, bnd_i, eng_i, pit_i, ton_i\}$. These predicted attributes are individually embedded and concatenated to form a dense word-level style token $\mathbf{c}_i$. To ensure precise control, we concatenate the semantic embedding of the current word, denoted as $\mathbf{e}(w_i)$, with the style token $\mathbf{c}_i$ to form a unified condition representation $\mathbf{v}_i = [\mathbf{e}(w_i) \oplus \mathbf{c}_i]$. Guided by this condition, the LLM autoregressively generates the corresponding speech chunk $\mathbf{s}_i$. Consequently, accounting for the initial silence segment $\mathbf{s}_0$ prior to speech onset, the complete sequence $\mathcal{S}$ for an $N$-word input is formulated as:
\begin{equation}
\mathcal{S} = [\mathbf{s}_0, \langle b \rangle, \mathbf{v}_1, \mathbf{s}_1, \langle b \rangle, \mathbf{v}_2, \mathbf{s}_2, \dots, \langle b \rangle, \mathbf{v}_N, \mathbf{s}_N]
\end{equation}

Crucially, this decoupled architecture inherently supports two distinct inference modes. In the \textit{free mode}, the model autonomously predicts $\mathcal{A}_i$, functioning as an intelligent prosodic planner to generate natural speech. In the \textit{control mode}, users can bypass the Word Decoder (implemented as a 2-layer MLP) and directly specify desired attribute values (e.g., forcing a high pitch or a specific tone for $w_i$). These user-specified values are seamlessly embedded to construct $\mathbf{c}_i$, thereby achieving zero-shot, highly deterministic word-level control without altering the underlying LLM weights.

\subsection{WordVoice-FM: Fine-Grained Style Modulation}

While WordVoice-LLM establishes prosodic planning, the generated speech tokens inherently suffer from acoustic degradation due to vector quantization. 
Inspired by the frame-level style modulation~\cite{hpro}, we introduce a fine-grained style modulation mechanism in the Flow Matching (FM) stage to ensure the generated speech strictly adheres to the designated attributes.
To bridge the resolution mismatch between the word-level style token $\mathbf{c}_i$ and frame-level acoustic features, we employ an upsampling strategy that strictly aligns with the LLM's sequence structure. Specifically, a learnable token $\mathbf{c}_0$ representing the initial silence segment $\mathbf{s}_0$ is upsampled by its duration, while each subsequent word $w_i$'s style token $\mathbf{c}_i$ is replicated frame-by-frame based on the sum of its active articulation and subsequent pause durations. This length regulation process effectively expands the discrete word-level styles into a frame-level sequence $\mathbf{C}_{f}$.

Subsequently, $\mathbf{C}_{f}$ is injected into the FM backbone as an explicit conditioning signal via a dedicated \textit{Word Style Modulation} module. Specifically, at each frame $t$, the aligned style token $\mathbf{c}_{f, t} \in \mathbf{C}_{f}$ is projected through a linear layer to generate the scale $\gamma_t$ and shift $\beta_t$ modulation parameters. The continuous embedding of the discrete speech token at frame $t$, denoted as $\mathbf{x}_t$, is then directly modulated as follows:
\begin{equation}
    \mathbf{\hat{x}}_t = \gamma_t \odot \mathrm{LayerNorm}(\mathbf{x}_t) + \beta_t
\end{equation}
where $\odot$ denotes element-wise multiplication. The modulated speech token representations $\mathbf{\hat{X}}$ are then fed into the subsequent FM layers. Through this fine-grained frame-level modulation, the generation process is explicitly guided by the precise energy dynamics, pitch contours, and temporal boundaries planned in the LLM stage. This effectively compensates for the acoustic details lost during quantization, ensuring that the final synthesized waveform exhibits highly accurate and decoupled word-level control.

\subsection{Training and Inference Strategy}

\textbf{Training Phase.} WordVoice-LLM is initialized with the Qwen2.5-0.5B~\cite{qwen2.5} backbone. During training, the model is optimized to autoregressively generate the speech tokens conditioned on the input text sequence $\mathcal{W}$ and the planned word-level styles. The standard autoregressive negative log-likelihood loss for generation is defined as:
\begin{equation}
    \mathcal{L}_{AR} = - \sum_{i=1}^{N} \log P(\mathbf{s}_i | \mathcal{W}, \mathcal{S}_{<s_i})
\end{equation}
where $\mathcal{S}_{<s_i}$ denotes the interleaved sequence generated prior to $\mathbf{s}_i$. Simultaneously, to enable the LLM to predict the acoustic attributes, we follow the pGSLM~\cite{pgslm} approach to discretize all continuous values, as standard LLMs typically struggle with continuous regression. Specifically, duration is quantized based on the 40ms frame rate, while pitch and energy are uniformly quantized into 20 discrete bins. Consequently, the prediction of the five attributes is formulated as classification tasks optimized via Cross-Entropy (CE) losses, denoted as $\mathcal{L}_{CE}^k$. To effectively balance these concurrent objectives alongside $\mathcal{L}_{AR}$, we employ an uncertainty-weighted multi-task loss~\cite{uwloss}:
\begin{equation}
    \mathcal{L}_{LLM} = \mathcal{L}_{AR} + \sum_{k \in \mathcal{A}} \left( \frac{1}{2\sigma_k^2} \mathcal{L}_{CE}^k + \log \sigma_k \right)
\end{equation}
where $\mathcal{A} = \{dur, bnd, eng, pit, ton\}$, and $\sigma_k$ represents the learnable observation noise parameter for each attribute task. 

\begin{table*}[t]
\centering
\caption{Subjective evaluation results on the Chinese and English test sets.}
\label{tab:subjective}
\resizebox{0.8\linewidth}{!}{
\begin{tabular}{lcccccc}
\toprule
\multirow{2}{*}{\textbf{Method}} & \multicolumn{3}{c}{\textbf{WordVoice-5A-zh-test}} & \multicolumn{3}{c}{\textbf{WordVoice-5A-en-test}} \\
\cmidrule(lr){2-4} \cmidrule(lr){5-7}
& \textbf{N-MOS} $\uparrow$ & \textbf{Spk-MOS} $\uparrow$ & \textbf{Ctrl-MOS} $\uparrow$ & \textbf{N-MOS} $\uparrow$ & \textbf{Spk-MOS} $\uparrow$ & \textbf{Ctrl-MOS} $\uparrow$ \\
\midrule
Mel-Recon & 3.961 $\pm$ 0.074 & 3.672 $\pm$ 0.083 & 4.074 $\pm$ 0.077 & 3.974 $\pm$ 0.077 & 3.884 $\pm$ 0.078 & 4.084 $\pm$ 0.076 \\
\midrule
CosyVoice3 & 3.550 $\pm$ 0.081 & \textbf{3.527 $\pm$ 0.088} & 3.028 $\pm$ 0.095 & 3.643 $\pm$ 0.086 & 3.838 $\pm$ 0.078 & 3.324 $\pm$ 0.094 \\
WordVoice-Free & 3.645 $\pm$ 0.083$^{*}$ & 3.326 $\pm$ 0.089 & 3.383 $\pm$ 0.092$^{**}$ & 3.725 $\pm$ 0.081 & \textbf{3.869 $\pm$ 0.078} & 3.512 $\pm$ 0.088$^{**}$ \\
WordVoice-Control & \textbf{3.689 $\pm$ 0.081}$^{**}$ & 3.324 $\pm$ 0.088 & \textbf{3.446 $\pm$ 0.092}$^{**}$ & \textbf{3.773 $\pm$ 0.082}$^{*}$ & 3.846 $\pm$ 0.077 & \textbf{3.645 $\pm$ 0.085}$^{**}$ \\
\bottomrule
\multicolumn{7}{@{}l}{\rule{0pt}{2.5ex}\footnotesize $^*$ and $^{**}$ denote statistically significant improvements over the CosyVoice3 baseline at $p < 0.05$ and $p < 0.01$, measured by the Mann-Whitney U test.}
\end{tabular}
}
\end{table*}

For WordVoice-FM, we adopt the standard Flow Matching objective. Let $x_1$ denote the ground-truth acoustic feature and $x_0$ denote the standard Gaussian noise. The model optimizes a velocity field $v_\theta$ conditioned on the modulated speech representations $\mathbf{\hat{X}}$. The loss function is defined as:
\begin{equation}
    \mathcal{L}_{FM} = \mathbf{E}_{t, x_0, x_1} \left[ \| v_\theta(x_t, \mathbf{\hat{X}}, t) - (x_1 - x_0) \|_2^2 \right]
\end{equation}
where $x_t$ is the interpolated state at time step $t \in [0,1]$. Furthermore, to enforce the model's reliance on the word-level style tokens, we randomly mask 30\% of the input speech tokens during training. This forces the FM model to reconstruct the missing acoustic details by heavily leveraging the provided word-level style conditions.

\begin{table*}[t]
\centering
\caption{Objective evaluation results on the Chinese and English test sets.}
\label{tab:objective}
\resizebox{\linewidth}{!}{
\begin{tabular}{lcccccccccccc}
\toprule
\multirow{2}{*}{\textbf{Method}} & \multicolumn{6}{c}{\textbf{WordVoice-5A-zh-test}} & \multicolumn{6}{c}{\textbf{WordVoice-5A-en-test}} \\
\cmidrule(lr){2-7} \cmidrule(lr){8-13}
& \textbf{WER} $\downarrow$ & \textbf{Dur-MAE} $\downarrow$ & \textbf{Eng-MAE} $\downarrow$ & \textbf{Pit-MAE} $\downarrow$ & \textbf{Bnd-ER} $\downarrow$ & \textbf{Ton-RER} $\downarrow$ & \textbf{WER} $\downarrow$ & \textbf{Dur-MAE} $\downarrow$ & \textbf{Eng-MAE} $\downarrow$ & \textbf{Pit-MAE} $\downarrow$ & \textbf{Bnd-ER} $\downarrow$ & \textbf{Ton-RER} $\downarrow$ \\
\midrule
Ground-Truth & - & 0.0142 & 0.0116 & 0.0094 & 5.20\% & 5.85\% & - & 0.0188 & 0.0091 & 0.0081 & 11.58\% & 7.01\% \\
Mel-Recon & 1.19\% & 0.0257 & 0.0266 & 0.0552 & 13.98\% & 13.90\% & 0.34\% & 0.0393 & 0.0247 & 0.0365 & 21.39\% & 16.20\% \\
\midrule
CosyVoice3 & \textbf{2.31\%} & 0.0549 & 0.1030 & 0.2030 & 32.47\% & 32.31\% & \textbf{1.06\%} & 0.0806 & 0.0899 & 0.1765 & 43.81\% & 40.92\% \\
WordVoice-Free & 2.58\% & 0.0500 & 0.0963 & 0.1855 & 31.33\% & 30.97\% & 1.31\% & 0.0696 & 0.0850 & 0.1568 & 42.37\% & 39.98\% \\
WordVoice-Control & 2.86\% & \textbf{0.0349} & \textbf{0.0486} & \textbf{0.1100} & \textbf{12.72\%} & \textbf{20.25\%} & 1.57\% & \textbf{0.0450} & \textbf{0.0475} & \textbf{0.0782} & \textbf{23.23\%} & \textbf{26.58\%} \\
\bottomrule
\end{tabular}
}
\end{table*}

\subsubsection{Inference.} During zero-shot inference, we construct the acoustic prompt by temporally aligning the reference audio and the text using MMS-FA~\cite{mmsfa}. We then extract the five acoustic attributes via our proposed data pipeline to form the condition sequence. For the target text generation, the LLM typically operates in the adaptive mode. However, users can actively intervene by specifying desired acoustic attributes for any specific words. These manual specifications directly replace the LLM's autonomous predictions to achieve active control. Finally, the discrete sequence generated by the LLM inherently provides explicit temporal boundaries. The FM module directly utilizes this LLM-generated duration information to perform the frame-level upsampling of the style tokens, ensuring perfect temporal alignment without external aligners.

\section{Experiments}
\label{sec:experiments}

\subsection{Experimental Setup}
We evaluate our method on the WordVoice-5A-test set, comprising about 2,000 Chinese and 1,500 English unseen utterances. For zero-shot synthesis, the audio corresponding to the first 30\% of the text in each utterance is extracted as the prompt audio to guide the generation of the remaining content. While recent SOTAs achieve high naturalness, they fundamentally lack explicit word-level control. Therefore, we select CosyVoice3 as a representative baseline to demonstrate that our method achieves unprecedented fine-grained control without compromising competitive zero-shot quality. We compare our method with four systems: 1) \textbf{Mel-Recon}: Waveforms reconstructed directly from Ground-Truth (GT) acoustic features using the FM vocoder (theoretical upper bound). 2) \textbf{CosyVoice3}: Baseline. 3) \textbf{WordVoice-Free}: Our model in adaptive prediction mode. 4) \textbf{WordVoice-Control}: Our model in manual intervention mode using GT attributes. Additionally, to evaluate the superiority of our explicit control mechanism, we introduce \textbf{MagicTTS} as a strong baseline for temporal attribute intervention. WeSCon is excluded from baselines because its abstract emotion control lacks deterministic mappings to specific acoustic attributes. For objective evaluation, synthesis stability is measured by Word Error Rate (WER), calculated using Qwen3-ASR-1.7B\footnote{\url{https://huggingface.co/Qwen/Qwen3-ASR-1.7B}} to compare the recognized texts of the generated audio against those of the GT audio. Control precision is quantified by the Mean Absolute Error (MAE) for continuous attributes (Dur, Eng, Pit) and the Error Rate (ER) for discrete attributes (Bnd, Ton). To avoid over-penalizing strict thresholds on continuous pitch, we introduce a relaxed Ton-ER (Ton-RER) that forgives misclassifications between adjacent categories (e.g., `Rise' vs. `Strong Rise'). To compute these objective metrics, we utilize Qwen3-FA to independently extract timestamps from the generated audio, which then serve as the basis for calculating the respective acoustic attributes. The proposed framework is trained using the Adam optimizer on 8 NVIDIA A800 GPUs, with the LLM optimized for 7 epochs and the FM for 20 epochs.

\subsection{Main Results}

\subsubsection{Subjective Evaluation.} 
We conducted crowd-sourced listening tests with 20 participants rating 80 utterances per model on a 5-point scale. We report Mean Opinion Scores (MOS)~\cite{mos} with 95\% confidence intervals for three metrics: Naturalness (N-MOS) for human-likeness, Speaker Similarity (Spk-MOS) for timbre consistency, and Word Style Controlled (Ctrl-MOS) for the precision of local prosodic manipulations.
As shown in Table~\ref{tab:subjective}, WordVoice-Control achieves the highest N-MOS and Ctrl-MOS. A Mann-Whitney U test confirms highly significant improvements in controllability (Ctrl-MOS, p$<$0.01) across all variants, alongside significant naturalness gains (N-MOS, p$<$0.05) for WordVoice-Control, demonstrating that explicit structural priors effectively guide the LLM to generate more expressive prosody.
Notably, CosyVoice3 retains a slightly higher Spk-MOS. This reflects an inherent trade-off: our paradigm injects strong localized variations that cause minor perturbations to global acoustic transitions—a worthwhile compromise for unprecedented word-level controllability.

\begin{figure}[t]
\begin{minipage}[hbtp]{1.0\linewidth}
    \centering
    \centerline{\includegraphics[width=0.95\linewidth]{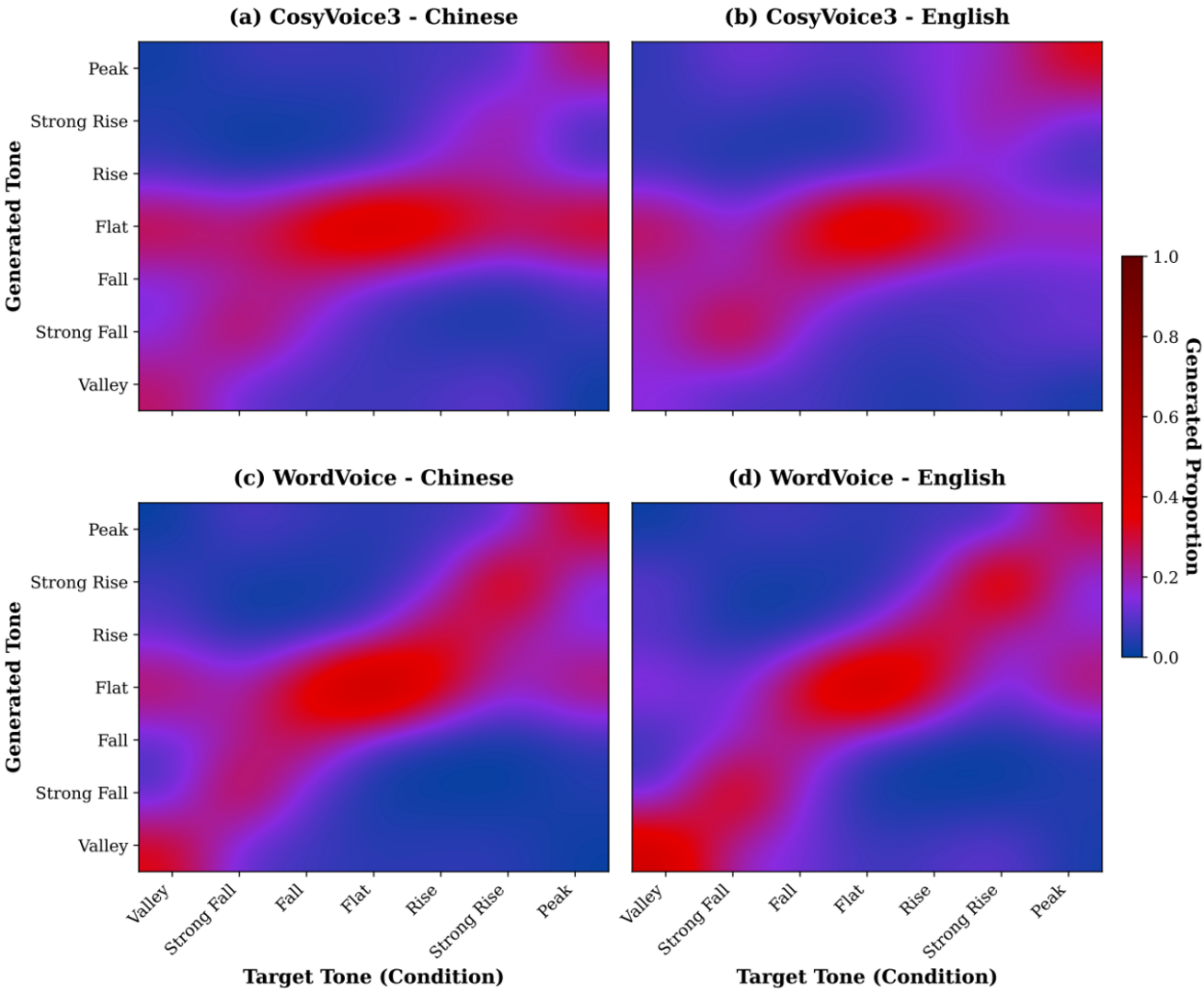}}
\end{minipage}
    \caption{2D density heatmaps of generated versus target category-specific tones.}
    \label{fig:tone_heatmap}
\end{figure}

\subsubsection{Objective Evaluation.} 
As shown in Table~\ref{tab:objective}, WordVoice-Free consistently outperforms CosyVoice3 across all acoustic metrics. This indicates that using word-level attributes as intermediate representations inherently guides the LLM to generate more natural, human-like prosody. In WordVoice-Control mode, all acoustic errors (MAE and ER) drop drastically, demonstrating highly precise manual intervention capabilities. Crucially, these objective improvements align with the Ctrl-MOS, verifying true perceptual control rather than mere pipeline bias. However, this structural conditioning introduces a minor trade-off: it slightly perturbs the LLM's linguistic robustness, causing a marginal increase in WER.

\begin{table}[t]
    \centering
    \caption{Comparison of explicit temporal control precision.}
    \label{tab:magictts_comp}
    \resizebox{\columnwidth}{!}{
    \begin{tabular}{lcccc}
    \toprule
    \multirow{2}{*}{\textbf{Method}} & \multicolumn{2}{c}{\textbf{WordVoice-5A-zh-test}} & \multicolumn{2}{c}{\textbf{WordVoice-5A-en-test}} \\
    \cmidrule(lr){2-3} \cmidrule(lr){4-5}
    & \textbf{Dur-MAE} $\downarrow$ & \textbf{Bnd-ER} $\downarrow$ & \textbf{Dur-MAE} $\downarrow$ & \textbf{Bnd-ER} $\downarrow$ \\
    \midrule
    MagicTTS-Dur & 0.0438 & - & 0.0593 & - \\
    MagicTTS-Pau & - & 16.10\% & - & 38.77\% \\
    \midrule
    \textbf{WordVoice} & \textbf{0.0383} & \textbf{12.82\%} & \textbf{0.0397} & \textbf{24.77\%} \\
    \bottomrule
    \end{tabular}
    }
\end{table}

\begin{table*}[t]
\centering
\caption{Objective evaluation results for decoupled single-attribute control and ablation studies.}
\label{tab:ablation_control}
\resizebox{\linewidth}{!}{
\begin{tabular}{lcccccccccc}
\toprule
\multirow{2}{*}{\textbf{Method}} & \multicolumn{5}{c}{\textbf{WordVoice-5A-zh-test}} & \multicolumn{5}{c}{\textbf{WordVoice-5A-en-test}} \\
\cmidrule(lr){2-6} \cmidrule(lr){7-11}
& \textbf{Dur-MAE} $\downarrow$ & \textbf{Eng-MAE} $\downarrow$ & \textbf{Pit-MAE} $\downarrow$ & \textbf{Bnd-ER} $\downarrow$ & \textbf{Ton-RER} $\downarrow$ & \textbf{Dur-MAE} $\downarrow$ & \textbf{Eng-MAE} $\downarrow$ & \textbf{Pit-MAE} $\downarrow$ & \textbf{Bnd-ER} $\downarrow$ & \textbf{Ton-RER} $\downarrow$ \\
\midrule
\multicolumn{11}{l}{\textit{Part 1: Decoupled Single-Attribute Control}} \\
\midrule
WordVoice (Dur-Only) & \textbf{0.0368} & 0.0947 & 0.1891 & 25.65\% & 29.83\% & \textbf{0.0452} & 0.0843 & 0.1588 & 40.97\% & 39.35\% \\
WordVoice (Eng-Only) & 0.0490 & \textbf{0.0498} & 0.1837 & 25.34\% & 29.69\% & 0.0693 & \textbf{0.0498} & 0.1545 & 40.84\% & 39.82\% \\
WordVoice (Pit-Only) & 0.0498 & 0.0931 & \textbf{0.1169} & 25.54\% & 27.74\% & 0.0691 & 0.0818 & \textbf{0.0899} & 40.34\% & 37.31\% \\
WordVoice (Bnd-Only) & 0.0488 & 0.0928 & 0.1860 & \textbf{13.22\%} & 29.67\% & 0.0683 & 0.0823 & 0.1561 & \textbf{23.01\%} & 39.48\% \\
WordVoice (Ton-Only) & 0.0497 & 0.0950 & 0.1807 & 25.37\% & \textbf{26.26\%} & 0.0691 & 0.0826 & 0.1532 & 40.12\% & \textbf{31.58\%}\\
\midrule
\multicolumn{11}{l}{\textit{Part 2: Ablation on WordVoice-FM}} \\
\midrule
Token-Recon (Orig-FM) & 0.0285 & 0.0796 & 0.1446 & 13.01\% & 22.77\% & 0.0349 & 0.0675 & 0.1084 & 23.24\% & 25.35\% \\
Token-Recon (WV-FM)   & \textbf{0.0284} & \textbf{0.0422} & \textbf{0.0958} & \textbf{11.38\%} & \textbf{20.47\%} & \textbf{0.0343} & \textbf{0.0392} & \textbf{0.0675} & \textbf{21.52\%} & \textbf{24.02\%} \\
\midrule
WordVoice (Orig-FM) & \textbf{0.0340} & 0.0861 & 0.1492 & 13.70\% & 25.50\% & \textbf{0.0442} & 0.0771 & 0.1161 & 23.85\% & 30.12\% \\
WordVoice (Full)   & 0.0349 & \textbf{0.0486} & \textbf{0.1100} & \textbf{12.72\%} & \textbf{20.25\%} & 0.0450 & \textbf{0.0475} & \textbf{0.0782} & \textbf{23.23\%} & \textbf{26.58\%} \\
\bottomrule
\end{tabular}
}
\end{table*}

\subsubsection{Category-Specific Tone Analysis.} 
To validate Ton-RER, Fig.~\ref{fig:tone_heatmap} visualizes the generated versus target tone distributions. Without explicit control, the baseline CosyVoice3 exhibits a relatively horizontal distribution, inherently defaulting to `Flat' prosody. In contrast, WordVoice-Control displays a pronounced diagonal alignment. Crucially, off-diagonal errors are mostly restricted to adjacent categories (e.g., `Fall' vs. `Strong Fall'). Since our 7-category system relies on hard discretization thresholds applied to continuous pitch contours, these adjacent shifts merely reflect minor natural acoustic fluctuations near these thresholds. This observation perfectly justifies the use of Ton-RER and visually confirms WordVoice's precise tone manipulation capability.

\subsubsection{Comparison with Word-level Control TTS.}
To further evaluate explicit temporal control, we compare WordVoice against MagicTTS on a 100-utterance subset. Since MagicTTS is architecturally limited to manipulating a single attribute per word, we evaluate its dedicated variants for duration (MagicTTS-Dur) and pause (MagicTTS-Pau) separately. As shown in Table~\ref{tab:magictts_comp}, WordVoice consistently achieves lower errors in both duration and boundary metrics across Chinese and English. This demonstrates that our proposed bound-token planning and fine-grained acoustic modulation enable more precise and stable temporal alignment, achieving superior control performance even when simultaneously manipulating multiple attributes within a unified framework.

\subsection{Decoupled Single-Attribute Control}
To verify the independence of our control mechanism, we evaluate WordVoice under single-attribute interventions. As shown in Part 1 of Table~\ref{tab:ablation_control}, the results exhibit an obvious diagonal phenomenon: constraining a specific attribute drastically reduces its corresponding error while having minimal impact on the others. This confirms that the word-level attributes are fundamentally decoupled. The only minor exception is tone. Unlike duration, boundary, energy, and pitch, which essentially serve as static global scalars for a word chunk, tone represents a dynamic variation contour. Due to this dynamic nature, tone is intrinsically more entangled with other acoustic features, making its absolute decoupling slightly more challenging.

\subsection{Ablation Studies on WordVoice-FM}
To evaluate the WordVoice-FM (\textit{WV-FM}) module, we conduct ablations under two settings as shown in Part 2 of Table~\ref{tab:ablation_control}: pure waveform reconstruction using ground-truth speech tokens (\textit{Token-Recon}), and the full WordVoice. The results reveal a clear division of modeling labor. Removing WV-FM (\textit{Orig-FM}) drastically degrades Eng-MAE and Pit-MAE in both settings, proving that WV-FM is essential for compensating continuous acoustic details lost during token quantization. Conversely, temporal attributes (duration and boundary) show almost no change, while tone exhibits only minor variations. This confirms that static temporal structures and dynamic variation traits are already effectively modeled within the LLM stage, relying minimally on downstream acoustic modulation.

\section{Conclusion}
\label{sec:conclusion}
In this paper, we present WordVoice to overcome the coarse-grained limitations of current LLM-based TTS, achieving precise, multi-dimensional word-level control. This advancement is driven by two core contributions: WordVoice-5A, a 4.7k-hour dataset with linguistically-guided acoustic annotations; and a dual-stage architecture integrating bound-token macro-prosodic planning with Flow Matching micro-acoustic modulation. Extensive experiments demonstrate that WordVoice enables highly decoupled manual intervention across five acoustic dimensions, delivering deterministic stylistic control while maintaining competitive synthesis stability. By transforming implicit speech generation into an explicit, interpretable process, this work opens broad possibilities for future research. Moving forward, we plan to integrate this mechanism into downstream tasks, such as providing interpretable generation for instruction-based TTS and serving as an acoustic chain-of-thought (CoT) for highly expressive spoken dialogue systems.

\begin{small}
\bibliographystyle{IEEEbib}
\bibliography{refs}
\end{small}

\end{document}